\documentclass[aps,prb,10pt]{revtex4-2}
\usepackage{subfig}
\usepackage{graphicx}
\usepackage{epsfig}
\usepackage{dcolumn}
\usepackage{bm}
\usepackage{times}
\usepackage{nicematrix}
\usepackage{amsmath}
\usepackage{color}
\def\beq{\begin{equation}}
\def\eeq{\end{equation}}
\def\beqn{\begin{align}}
\def\eeqn{\end{align}}

\begin{document}
\title{Comment on ``Standard and non-standard Lagrangians for dissipative dynamical systems with variable coefficientes"}
\author{Gabriel Gonz\'alez}
\affiliation{C\'atedra CONCAYT--Universidad Aut\'onoma de San Luis Potos\'i, San Luis Potos\'i, 78000 MEXICO}
\affiliation{Coordinaci\'on para la Innovaci\'on y la Aplicaci\'on de la Ciencia y la Tecnolog\'ia, Universidad Aut\'onoma de San Luis Potos\'i,San Luis Potos\'i, 78000 MEXICO}


\begin{abstract}
Z.E. Musielak has reported in 2008 J. Phys. A: Math. Theor. {\bf 41} 055205 methods to obtain standard and non-standard Lagrangians and identify classes of equations of motion that admit a Lagrangian description. In this comment we show how to obtain new non-standard Lagrangians using the non-standard Lagrangians previously found. In particular, it is demonstrated that for every non-standard Lagrangian one can generate a new non-standard Lagrangian associated to a new equation of motion.
\end{abstract}

\maketitle
Lagrangians are very useful because they can be used to formulate classical and quantum theories, and can also be used to find a conservation law or to estimate the solution of a differential equation.\cite{musielak,jan} \\
Standard Lagrangians are quadratic forms with respect to $\dot{x}$, for the one dimensional case the standard Lagrangian takes the form
\begin{equation}\label{eq01}
  \mathcal{L}(x,\dot{x},t)=\frac{1}{2}P(x,t)\dot{x}^2+Q(x,t)\dot{x}+R(x,t)
\end{equation}
With a given Lagrangian we can obtain the equations of motion of the system by using the Euler-Lagrange equations
\begin{equation}\label{eq02}
  \frac{d}{dt}\left(\frac{\partial \mathcal{L}}{\partial\dot{x}}\right)-\frac{\partial\mathcal{L}}{\partial x}=0
\end{equation}
Expanding the differentiation in equation (\ref{eq02}) we get
\begin{equation}\label{eq03}
  \ddot{x}\frac{\partial^2 \mathcal{L}}{\partial\dot{x}^2}+\dot{x}\frac{\partial^2 \mathcal{L}}{\partial\dot{x}\partial x}+\frac{\partial^2 \mathcal{L}}{\partial\dot{x}\partial t}-\frac{\partial\mathcal{L}}{\partial x}=0
\end{equation}
If we differentiate equation (\ref{eq03}) with respect to $\dot{x}$ we get the following equation\cite{glez}
\begin{equation}\label{eq04}
  \frac{\partial}{\partial\dot{x}}\left(\ddot{x}M\right)+x\frac{\partial M}{\partial x}+\frac{\partial M}{\partial t}=0
\end{equation}
where $M(x,\dot{x},t)=\partial^2 \mathcal{L}/\partial\dot{x}^2$. It is important to note that for a standard Lagrangian $\partial^2 \mathcal{L}/\partial\dot{x}^2=M(x,,t)$, i.e. it does not depend explicitly on $\dot{x}$, therefore if a non-standard Lagrangian exists then the following condition most hold true, i.e.
\begin{equation}\label{eq04a}
\frac{\partial M}{\partial \dot{x}}\neq0
\end{equation}
Suppose now that we have the following equation of motion
\begin{equation}\label{eq05}
  \ddot{x}=f_0(x,\dot{x},t)-\frac{g(x,t)}{M(x,\dot{x},t)}
\end{equation}
If we substitute equation (\ref{eq05}) into equation (\ref{eq04}) we obtain
\begin{equation}\label{eq06}
  \frac{\partial}{\partial\dot{x}}\left(f_0M\right)+x\frac{\partial M}{\partial x}+\frac{\partial M}{\partial t}=0
\end{equation}
Equation (\ref{eq06}) tells us that if we know the non standard Lagrangian $\mathcal{L}_0$ associated with the equation of motion given by
\begin{equation}\label{eq07}
  \ddot{x}_0=f_0(x,\dot{x},t)
\end{equation}
then the non-standard Lagrangian associated with equation (\ref{eq05}) can be constructed by partially integrating $\partial^2 \mathcal{L}_0/\partial\dot{x}^2$, i.e. 
\begin{equation}\label{eq08}
\mathcal{L}(x,\dot{x},t)=\int\int \frac{\partial^2 \mathcal{L}_0}{\partial\dot{x}^2}d\dot{x}d\dot{x}+Q(x,t)\dot{x}+R(x,t)
\end{equation}
{\bf Proposition} Suppose $\mathcal{L}_0$ is a non-standard Lagrangian for $\ddot{x}_0$; then there exists a non-standard Lagrangian given by
\begin{equation}\label{eqt}
\mathcal{L}(x,\dot{x},t)=\mathcal{L}_0(x,\dot{x},t)-\int g(x,t)dx
\end{equation}
which describes the following equation of motion
\begin{equation}\label{eqt1}
\ddot{x}=\ddot{x}_0-\frac{g(x,t)}{\frac{\partial^2 \mathcal{L}_0}{\partial\dot{x}^2}}
\end{equation}
{\bf Proof} We substitute the Lagrangian of equation (\ref{eqt}) into the Euler-Lagrange equation (see equation (\ref{eq02})) and after using equation (\ref{eqt1}) and the fact that $\mathcal{L}_0$ describes the equation of motion $\ddot{x}_0$ we get an identity which validates the proposition.\\

Let us now work out an example, the non-standard Lagrangian associated with the standard harmonic oscillator, i.e. $\ddot{x}=-\omega^2x$, is given by\cite{havas}
\begin{equation}\label{eq09}
  \mathcal{L}_0(x,\dot{x},t)=\frac{\dot{x}}{\omega x}\arctan\left(\frac{\dot{x}}{\omega x}\right)-\frac{1}{2}\ln\left(\dot{x}^2+\omega^2x^2\right)
\end{equation}
Using equation (\ref{eq09}) we have
\begin{equation}\label{eq10}
  \frac{\partial^2 \mathcal{L}_0}{\partial\dot{x}^2}=\frac{1}{\omega^2x^2+\dot{x}^2}
\end{equation}
Now, suppose we want to obtain the non-standard Lagrangian of the following equation of motion
\begin{equation}\label{eq11}
  \ddot{x}=-\omega^2x-x\left( \omega^2x^2+\dot{x}^2\right)
\end{equation}
Equation (\ref{eq11}) is of the Li\'enard type non-linear oscillator which shows very unusual properties\cite{lienard} and corresponds to the form of equation (\ref{eqt1}) by taking $\ddot{x}_0=-\omega^2x$, $g(x,t)=x$ and $\partial^2 \mathcal{L}_0/\partial\dot{x}^2=\left( \omega^2x^2+\dot{x}^2\right)^{-1}$. Using equation (\ref{eqt}) we can obtain the following non-standard Lagrangian
\begin{equation}\label{eq12}
\mathcal{L}(x,\dot{x},t)=\frac{\dot{x}}{\omega x}\arctan\left(\frac{\dot{x}}{\omega x}\right)-\frac{1}{2}\ln\left(\dot{x}^2+\omega^2x^2\right)-\frac{x^2}{2}
\end{equation}
One can use this approach to generalized the non-standard Lagrangians obtained by Z.E. Musielak, for example the equation of motion given by
\begin{equation}\label{eq13}
\ddot{x}+B(t)\dot{x}+\frac{2}{3}\left(\dot{B}(t)+\frac{1}{3}B(t)^2\right)x+g(x,t)\left(\dot{x}+\frac{2}{3}B(t)x\right)^3=0
\end{equation}
admits the following non-standard Lagrangian
\begin{equation}\label{eq14}
\mathcal{L}(x,\dot{x},t)=\frac{1}{\dot{x}+\frac{2}{3}B(t)x}-\int g(x,t)dx
\end{equation}
If $g(x,t)=0$ then que recover the non-standard Lagrangian obtained by Musielak.\cite{musielak}\\
In conclusion, once a non-standard Lagrangian has been found for a given equation of motion, then it is possible to generate another non-standard Lagrangian for a new equation of motion. A new result is that the
forms of equations with the new non-standard Lagrangian have linear, quadratic and {\it cubic} dissipative terms.
\section*{Acknowledgments}

I would like to acknowledge support by the program Cátedras Conacyt through project 1757 and from project A1-S-43579 of SEP-CONACYT Ciencia Básica and Laboratorio Nacional de Ciencia y Tecnología de Terahertz.

\end{document}